\documentclass[preprint,12pt]{elsarticle}

\usepackage{lineno}
\usepackage{amssymb}
\usepackage{amsmath}
\usepackage{mhchem}
\usepackage{url} 
\usepackage{orcidlink}

\journal{NIM A}

\begin{document}

\begin{frontmatter}



\title{
Development of Deep Neural Network First-Level Hardware Track Trigger for the Belle II Experiment} 


\author[1]{Y.-X.~Liu\,\orcidlink{0000-0002-8374-3947}}
\author[1,2]{T.~Koga\,\orcidlink{0000-0002-1644-2001}}
\author[1]{H.~Bae\,\orcidlink{0000-0003-1393-8631}}
\author[12]{Y.~Yang\,\orcidlink{0000-0001-7275-3982}}
\author[4]{C.~Kiesling\,\orcidlink{0000-0002-2209-535X}}
\author[5]{F.~Meggendorfer\,\orcidlink{0000-0002-1466-7207}}
\author[5]{K.~Unger\,\orcidlink{0000-0001-7378-6671}}
\author[4]{S.~Hiesl}
\author[4]{T.~Forsthofer}
\author[1,2]{A.~Ishikawa\,\orcidlink{0000-0002-3561-5633}}
\author[10]{Y.~Ahn\,\orcidlink{0000-0001-6820-0576}}
\author[5]{T.~Ferber\,\orcidlink{0000-0002-6849-0427}}
\author[5]{I.~Haide\,\orcidlink{0000-0003-0962-6344}}
\author[5]{G.~Heine\,\orcidlink{0009-0009-1827-2008}}
\author[3]{C.-L.~Hsu\,\orcidlink{0000-0002-1641-430X}}
\author[3]{A.~Little\,\orcidlink{0009-0008-4974-3661}}
\author[6]{H.~Nakazawa\,\orcidlink{0000-0003-1684-6628}}
\author[5]{M.~Neu\,\orcidlink{0000-0002-4564-8009}}
\author[5]{L.~Reuter\,\orcidlink{0000-0002-5930-6237}}
\author[8]{V.~Savinov\,\orcidlink{0000-0002-9184-2830}}
\author[9]{Y.~Unno\,\orcidlink{0000-0003-3355-765X}}
\author[7]{J.~Yuan\,\orcidlink{0009-0005-0799-1630}}
\author[11]{Z.~Xu\,\orcidlink{0009-0005-1048-4744}}
\affiliation[1]{organization={SOKENDAI (The Graduate University for Advanced Studies)},
            city={Hayama},
            postcode={240-0193}, 
            }

\affiliation[2]{organization={High Energy Accelerator Research Organization (KEK)},
            city={Tsukuba},
            postcode={305-0801}, 
            }

\affiliation[12]{organization={Fudan University},
            city={Shanghai},
            postcode={200433}}
\affiliation[4]{organization={Max-Planck-Institut für Physik},
            city={München},
            postcode={80805}
            }
\affiliation[5]{organization={Karlsruher Institut für Technologie~(KIT)},
            city={Karlsruhe},
            postcode={76131}}
\affiliation[10]{organization={Korea University},
            city={Seoul},
            postcode={02841}}
            
\affiliation[3]{organization={School of Physics, University of Sydney},
            city={Sydney},
            postcode={2006}}
\affiliation[6]{organization={Department of Physics, National Taiwan University},
                city={Taipei},
                postcode={10617}, 
}
\affiliation[8]{organization={University of Pittsburgh},
            city={Pittsburgh},
            postcode={15260}}

\affiliation[9]{organization={Department of Physics and Institute of Natural Sciences, Hanyang University},
            city={Seoul},
            postcode={04763}, 
            }
\affiliation[7]{organization={Jilin University},
        city={Jilin},
        postcode={130012}}

\affiliation[11]{organization={The University of Tokyo},
            city={Tokyo},
            postcode={113-8654}}

\begin{abstract}
The Belle~II experiment at the SuperKEKB accelerator is designed to explore physics beyond the Standard Model with unprecedented luminosity. As the beam intensity increased, the experiment faced significant challenges due to higher beam-induced background, leading to a high trigger rate and placing limitations on further luminosity increases. To address this problem, we developed trigger logic for tracking using deep neural network (DNN) technology on an FPGA for the Belle II hardware trigger system, employing high-level synthesis techniques. By leveraging drift time and hit pattern information from the Central Drift Chamber and incorporating a simplified self-attention architecture, the DNN track trigger significantly improves track reconstruction performance at the hardware level. Compared to the existing neural track trigger, our implementation reduces the total track trigger rate by 37\% while improving average efficiency for the signal tracks from 96\% to 98\% for charged tracks with transverse momentum $> 0.3$\,GeV. This upgrade ensures the long-term viability of the Belle~II data acquisition system as luminosity continues to increase.

\end{abstract}




\end{frontmatter}



\section{Introduction}

The Belle II Experiment~\cite{belleIITDR}, located at the asymmetric 
7~GeV electron - 4~GeV positron 
collider SuperKEKB~\cite{SKEKB} in Tsukuba, Japan, has been in operation since 2019. It aims to accumulate an integrated luminosity of 50~$\text{ab}^{-1}$ with a peak luminosity of $6 \times 10^{35} \, \text{cm}^{-2} \, \text{s}^{-1}$ at a center-of-mass energy of 10.58 GeV. This corresponds to the $\Upsilon(4S)$ resonance, which decays into a pair of $B$ mesons. The primary physics goals of Belle II are to explore new physics in the flavor sector at the intensity frontier and to enhance the precision of measurements for Standard Model parameters~\cite{Physicsbook}.

Belle II is a general-purpose detector consisting of seven sub-detectors and a superconducting solenoid, arranged cylindrically around the $e^+ e^-$ beam interaction point (IP). Moving outward from the IP, the Belle II detector consists of the Pixel Vertex Detector (PXD), Silicon Vertex Detector (SVD), Central Drift Chamber (CDC), Time-Of-Propagation detector (TOP), Aerogel Ring-Imaging Cherenkov detector (ARICH), Electromagnetic Calorimeter (ECL), and the $K_L$ and Muon detector (KLM). In this paper, we use a right-handed coordinate system with the origin at the IP, the $z$-axis aligned with the solenoid axis (approximately in the direction of the electron beam), and the polar angle defined with respect to $\hat{z}$. The azimuthal angle is measured relative to the direction pointing toward the inside of the accelerator ring.

At the designed instantaneous luminosity, the expected interesting events, including $\Upsilon(4S) \to B\bar{B}$; $\mu^+\mu^-$, $\tau^+\tau^-$, $\gamma\gamma$ and continuum hadron production by $e^+e^-$ annihilation; and prescaled $e^+e^-\to e^+ e^-$ scattering, occur at a rate of approximately 15\,kHz~\cite{belleIITDR}, 
while the major beam background induced by 
beam-gas interaction and Touscheck scattering~\cite{beambkg} can reach rates on the order of a few MHz.
To manage this, a first-level (L1) trigger system is employed to retain interesting events while rejecting most beam backgrounds, thereby reducing the data throughput to 
the Data Acquisition System (DAQ)~\cite{DAQ} 
with the maximum trigger rate of 30\,kHz. 
In Belle II, the L1 trigger is primarily implemented using Field Programmable Gate Arrays (FPGAs) and collects inputs from sub-detectors' first-in-first-out (FIFO) buffers. It is a hard-wired, deadtime-free system. To avoid FIFO overflow, a strict real-time deadline of 5\,µs is defined for the L1 trigger. The Belle II L1 trigger system consists of four components: the CDC, the ECL, the TOP, and the KLM trigger systems~\cite{L1TRG}.
Their signals are sent to Global Reconstruction Logic (GRL) for track-cluster matching~\cite{GRL} and to Global Decision Logic (GDL) to make the final L1 trigger decision.  
The entire trigger system operates with a common 127.216\,MHz system clock (corresponding to a cycle time of 7.8\,{ns}), which is derived by dividing the SuperKEKB RF reference clock by four.

Most final-state particles from physics events of interest originate from a small collision volume around the IP, except for the decay products of long-lived particles, 
such as $K_S^0$ or $\Lambda^0$. However, major beam background particles from beam-gas interaction and Touschek scattering can enter the Belle II detector and mimic desired annihilation events. These background events typically have large displacement vertices from the IP and should be removed by the CDC trigger system, which tracks charged particles and fits the track to extract the track parameters. The current CDC trigger relies on the track momentum and $z_0$ of the track starting point, the latter predicted using a Multilayer Perceptron (MLP) with a single hidden layer (hereafter referred to as the ``baseline'') to distinguish between beam background and interesting events~\cite{Neural}. 
However, under beam background conditions experienced while the luminosity was increasing in late 2022 physics data-taking, the baseline trigger system exhibited a 
high trigger rate of a few kHz. 
This was primarily due to limited resolution of the track vertex, which led to background tracks being misclassified as having small $|z_0|$. 
If the beam background increases further with higher luminosity, the trigger rate may exceed the limitation of the 30\,kHz in the future.

To address this issue, we report on the Belle II L1 CDC track trigger upgrade using a simplified attention architecture with a fully connected classifier, enriched input features, and an upgraded FPGA board. The attention mechanism, originally introduced in the context of natural language processing by Vaswani et al.~\cite{vaswani2017attention}, enables models to dynamically focus on the most relevant parts of the input. In our implementation, we adopt a simplified version of this architecture to enhance track feature extraction in the presence of high background rates. This is the first application of an attention-based Deep Neural Network (DNN) in the hardware trigger system for collider experiments.

The remainder of this paper is organized as follows. 
In Section~\ref{sec:cdctrg}, we describe the Belle II CDC trigger system. Section~\ref{sec:dnn} details the development, training, and tuning of the DNN track trigger. The firmware implementation workflow is presented in Section~\ref{sec:firm}. Section~\ref{sec:perf} evaluates the system performance based on Belle II physics data. Finally, we provide a summary in Section~\ref{sec:summary}.

\section{Belle II CDC Trigger System} \label{sec:cdctrg}
\begin{figure}[t]
\centering
\includegraphics[width=0.9\textwidth]{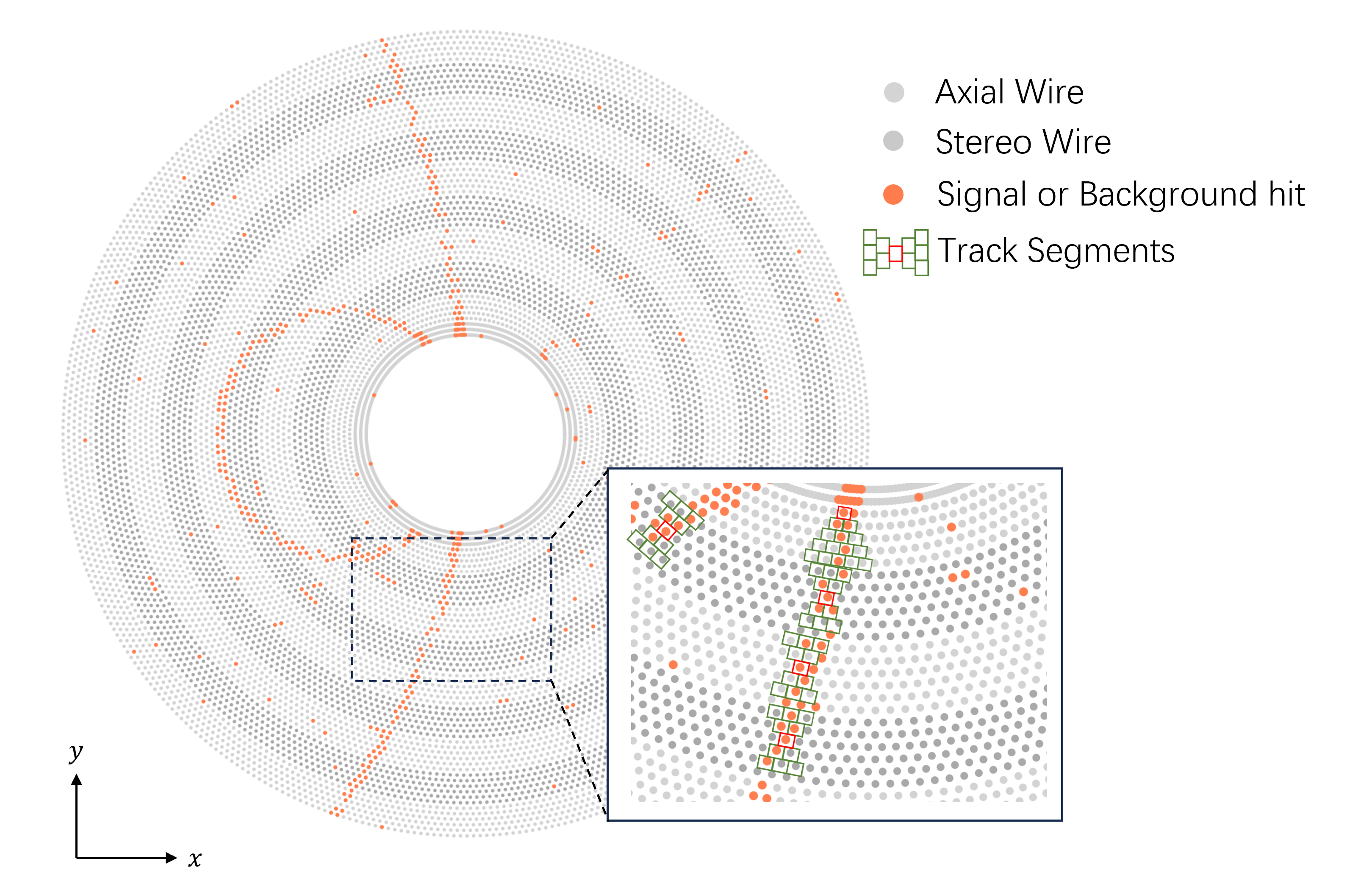}
\caption{Schematic view of the CDC cross section in the $r$-$\phi$ plane. Light gray dots correspond to axial sense wires, and dark gray dots are stereo sense wires. Every 8 or 6 layers of axial or stereo wires make a Super layer (SL). Orange dots correspond to the wires with hits. The magnified section shows the Track Segments built using specific patterns of hit wires. The track segments are the basic units for trigger logic.}\label{fig:crosssection}
\end{figure}

The Central Drift Chamber (CDC)~\cite{CDC} of the Belle II detector is a cylindrical wire chamber with an outer radius of 113\,cm and an inner radius of 16\,cm. It comprises 14,366 sense wires and 42,240 field wires arranged in 56 layers, grouped into 9 super layers (SLs). The cross section of CDC is shown in Fig.~\ref{fig:crosssection}. The innermost SL consists of 8 layers to mitigate beam-induced backgrounds, while the remaining SLs contain 6 layers each. The CDC operates with a 50\% He and 50\% \ce{C2H6} gas mixture. The wire directions for each of the nine SLs alternate between axial wires, which are parallel to the beam axis, and stereo wires, which are skewed by $67.4$\,mrad to $74.9$\,mrad in the positive direction and $-58.6$\,mrad to $-79.4$\,mrad in the negative direction. Axial wires enable 2D track reconstruction in the $r$-$\phi$ plane transverse to the beam direction, while stereo wires provide 3D spatial resolution.
CDC provides hit timing with 1\,ns resolution (TDC), integrated charge (ADC) and time-over-threshold for every wire for the offline analysis. However, at L1 trigger level, only wires in the inner 5 layers for SL\,1-8 and outer 5 layers for SL\,0 are available. We only have 2\,ns resolution TDC for priority wires. Besides that, 32\,ns resolution TDC and a flag to indicate whether the ADC passes the threshold are also available for every wire.

\begin{figure}[t]
\centering
\includegraphics[width=0.89\textwidth]{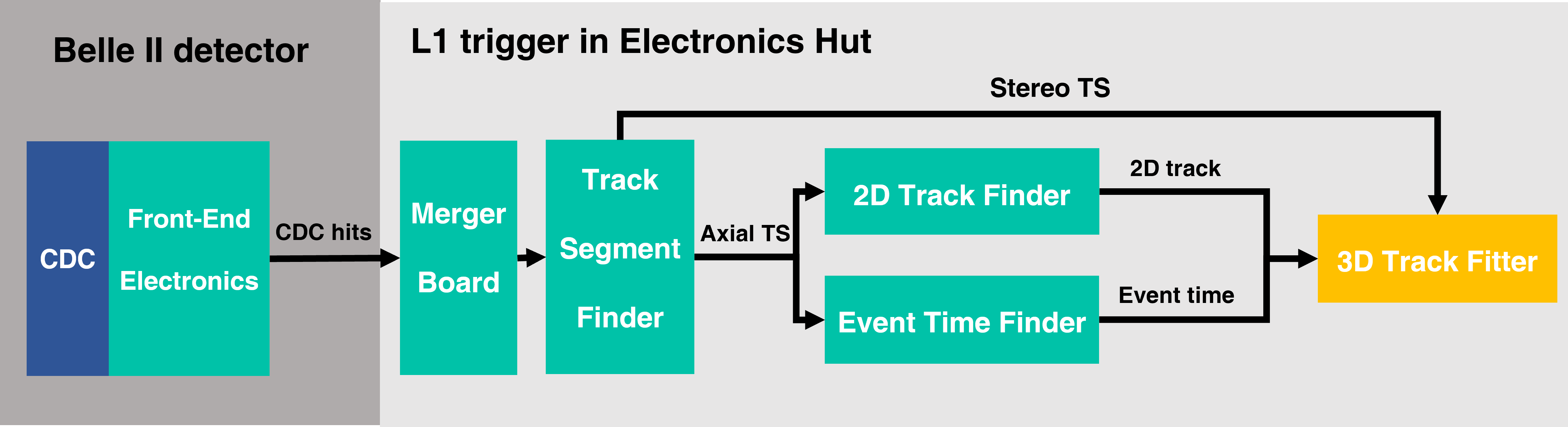}
\caption{
Schematic of the Belle~II L1 CDC trigger system. 
It collects raw CDC hits from CDC 
FEE, 
builds Track Segments (TSs), 
determines the event time, 
and processes TSs 
to find 2D tracks 
and 
to fit 3D tracks. }\label{fig:cdcworkflow}

\end{figure}

\begin{figure}[t]

\centering
\includegraphics[width=0.89\textwidth]{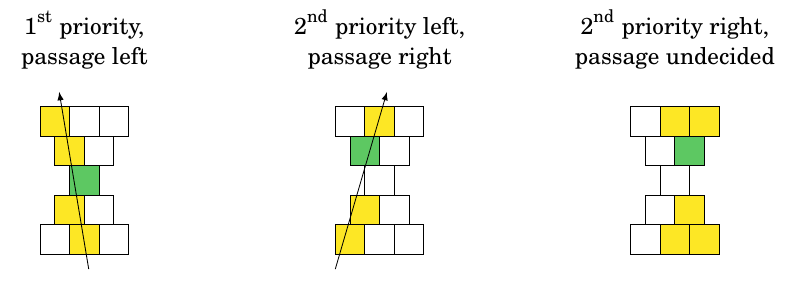}
\caption{Examples of the standard Track Segments patterns. Each cell corresponds to one CDC wire. Green cells are the priority hit wires, and yellow cells are hit wires. Based on the hit pattern, we define passage direction for tracks.\cite{sara}}\label{fig:TSF}

\end{figure}

The CDC trigger workflow is shown in Fig.~\ref{fig:cdcworkflow}. The raw CDC wire hits 
and timing 
information from CDC front-end electronics (FEE)~\cite{CDCFE} is processed from the merger board to the Track Segment Finder (TSF)~\cite{cdctrgcom}. TSF forms Track Segments (TSs) from the raw CDC hits with specific patterns as shown in Fig.~\ref{fig:TSF}. TSs are used as elements for following CDC trigger workflow to compress the data size and suppress the noise. 

After TSF, the axial TSs are fed into the 2D track finder and event time finder (ETF)~\cite{etf}. The 2D track finder constructs 2D tracks in the $r$-$\phi$ plane using Hough transformation, while the ETF determines the event timing ($t_0$) for each 2D track. Combined with the 2D track, $t_0$, and stereo TSs, the 3D track fitting is performed 
to estimate the origin of the tracks $z_0$ as well as their polar angle $\theta$. Both 2D and 3D tracks are transmitted to the GRL.

The Merger modules are composed of specially made boards with Altera Arria II FPGA. The remaining modules utilize customized Belle~II Universal Trigger (UT) Boards of the 3rd and 4th generations (UT3 and UT4). Table~\ref{tab:ut} lists specifications of the UT boards. The TSF, 2D track finder, ETF, and Neural-Network trigger module consist of nine UT4, four UT4, one UT4, and four UT3 boards, respectively. The scale of current Neural-Network trigger is constrained by the resource limitations of the UT3.

\begin{table}[t]
\centering
\caption{Specifications of Belle~II universal trigger boards.}
\scalebox{0.9}{
\begin{tabular}{l | c | c}
\hline
Generation & UT3  & UT4 \\
\hline
FPGA family  & AMD Virtex 6   & AMD Virtex Ultrascale \\
FPGA type    & XC6VHX380/565T & XCVU080/160/190       \\
The number of logic cells (k) & 380/565 & 975/2027/2350 \\
DSP slices   &   864/864      & 672/1560/1800  \\
Optical port & 5\,Gbps  GTX 40 lanes & 16\,Gbps GTH 32 lanes \\
             & 10\,Gbps GTH 24 lanes & 25\,Gbps GTY 32 lanes \\
The number of LVDS port & 128 & 64 \\
RAM          & -- & DDR4 32GiB \\
Sub FPGA     & -- & AMD Artix XC7A15T \\
\hline
\end{tabular}}
\label{tab:ut}
\end{table}


\section{DNN Track Trigger} \label{sec:dnn}

We upgraded the 3D track fitting module by implementing a DNN on the UT4 board, hereafter called the DNN track trigger. This upgrade enhances the baseline model by integrating a simplified self-attention architecture, allowing for more effective utilization of additional information from the TSF, which includes the wire timing for every wire in the TSF with 32\,ns resoultion. This extra timing information is newly transmitted via the UT4 board and not used in the baseline model. This section describes the working principles and model architecture of the DNN track trigger.

\subsection{Tracking Principle}
Charged particle tracking consists of two steps:
\begin{enumerate}
\item Track finding: identifying TSs 
associated with 
the same track.
\item Track fitting: determining track parameters ($\phi_0$, $\omega$, $z_0$, and $\theta_0$).
\end{enumerate}
 In the Belle II trigger system, at the 3D tracking stage, we already find a 2D track in the $r$-$\phi$ plane with related axial TSs. Thus, we only perform track finding for stereo TSs. Following the approach chosen in the baseline model, we select the best stereo TSs from each SL within a predefined $\Delta\phi$ range of the 2D track in the $r$-$\phi$ plane. If multiple TSs are found, only the one with the shortest drift time  for priority wire and known drift direction is used. A track is considered a valid 3D track only if at least 3 out of 4 SLs have selected stereo TSs.

After track finding, we perform track fitting using all the stereo and axial TSs. A charged particle track in a constant magnetic field follows a helical trajectory. Since the 2D track finder assumes tracks originate from the interaction point in the $r$-$\phi$ plane, the helical trajectory can be expressed as:
\begin{equation}
\begin{pmatrix} x(\mu) \\ y(\mu) \\ z(\mu) \end{pmatrix} =
\begin{pmatrix}
r\left[\sin\left(\frac{\mu}{r} - \phi_0\right) + \sin\phi_0\right] \\
r\left[\cos\left(\frac{\mu}{r} - \phi_0\right) - \cos\phi_0\right] \\
\cot\theta_0 \cdot \mu + z_0
\end{pmatrix},
\label{equ.main}
\end{equation}
where $\mu$ is the arc length of the transverse track projection. Since $\phi_0$ and $r$ are provided by the 2D track finder, the DNN track trigger, as in the baseline model, focuses on fitting $z_0$ and $\theta_0$.

\begin{figure}[t]
\centering
\includegraphics[width=0.87\textwidth]{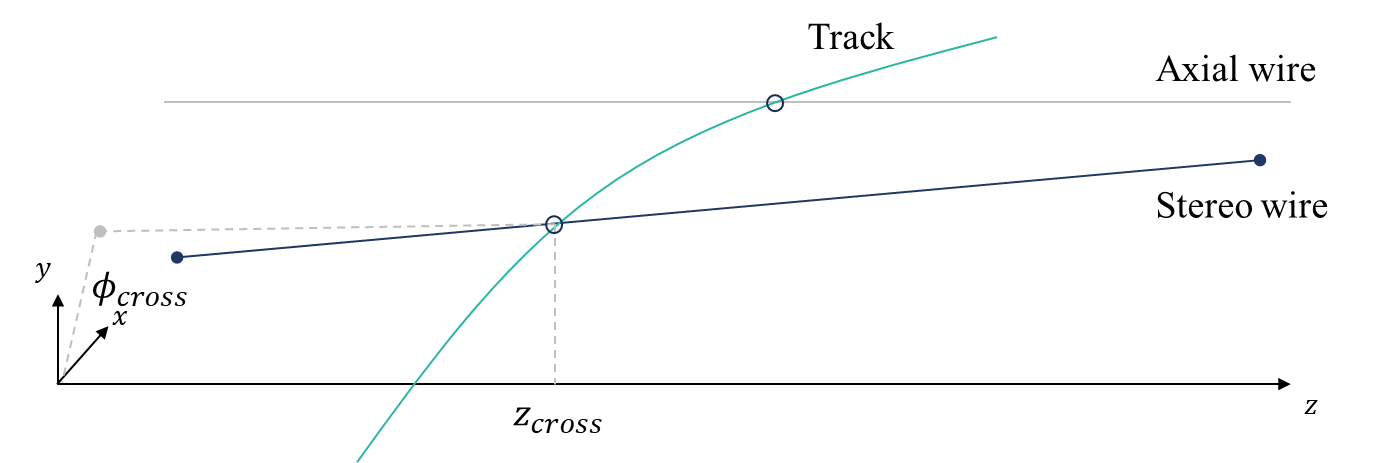}
\caption{Schematic of charge track hits on the stereo and axial wire. }\label{fig:principle}
\end{figure}

As shown in Fig.~\ref{fig:principle}, a linear approximation of the stereo wire in the $z$-$\phi$ plane allows us to express the charged particle crossing point $z_{\text{cross}}$ as:
\begin{equation}
z_{\text{cross}} = \frac{(z_F - z_B) \cdot (\phi_{\text{cross}} - \phi_B)}{\phi_F - \phi_B} - z_B,
\end{equation}
where the indices $F$ and $B$ denote the forward and backward endplates, and $z_B, z_F, \phi_B, \phi_F$ are constants specific to each stereo wire. The parameter $\phi_{\text{cross}}$ represents the crossing point of the 2D track and the stereo wire projection.
Taking drift time into account, the charged track does not exactly cross the wire but instead hits a point offset from $\phi_{\text{cross}}$. This hit position can be approximated as:
\begin{equation}
\phi_{\text{hit}} = \phi_{\text{cross}} \pm \arcsin\left(\frac{v_{\text{drift}} \cdot t_{\text{drift}}}{r_{\text{wire}}}\right) \approx \phi_{\text{cross}} \pm \frac{v_{\text{drift}} \cdot t_{\text{drift}}}{r_{\text{wire}}},
\end{equation}
where $t_{\text{drift}}$ and $v_{\text{drift}}$ denote the drift time and drift velocity, respectively. The sign corresponds to the drift direction, which can be determined from TS patterns. Additionally, the arc length of the crossing point, $\mu_{\text{cross}}$, can be derived from the 2D track and stereo wire geometry. With more than two hit points $(z_{\text{cross}}, \mu_{\text{cross}})$, we can fit the 3D track and extract $(z_0, \theta_0)$.

Using the above parameters for the selected stereo TSs, we can fit the 3D track parameters in the $\mu$-$z$ plane by minimizing the $\chi^2$ between the selected TSs and the fitted tracks. However, due to the FPGA resource limitation and the presence of massive background hits, it is quite hard to fit the track with limited latency. 
Therefore, a neural network-based approach had been chosen for the track trigger. Here we present a more sophisticated network architecture compared to the baseline model for improved accuracy of estimated track parameters and tracking robustness under different background conditions.

\subsection{DNN architecture and hyperparameter tuning}
\begin{figure}[t]
\centering
\includegraphics[width=0.45\textwidth]{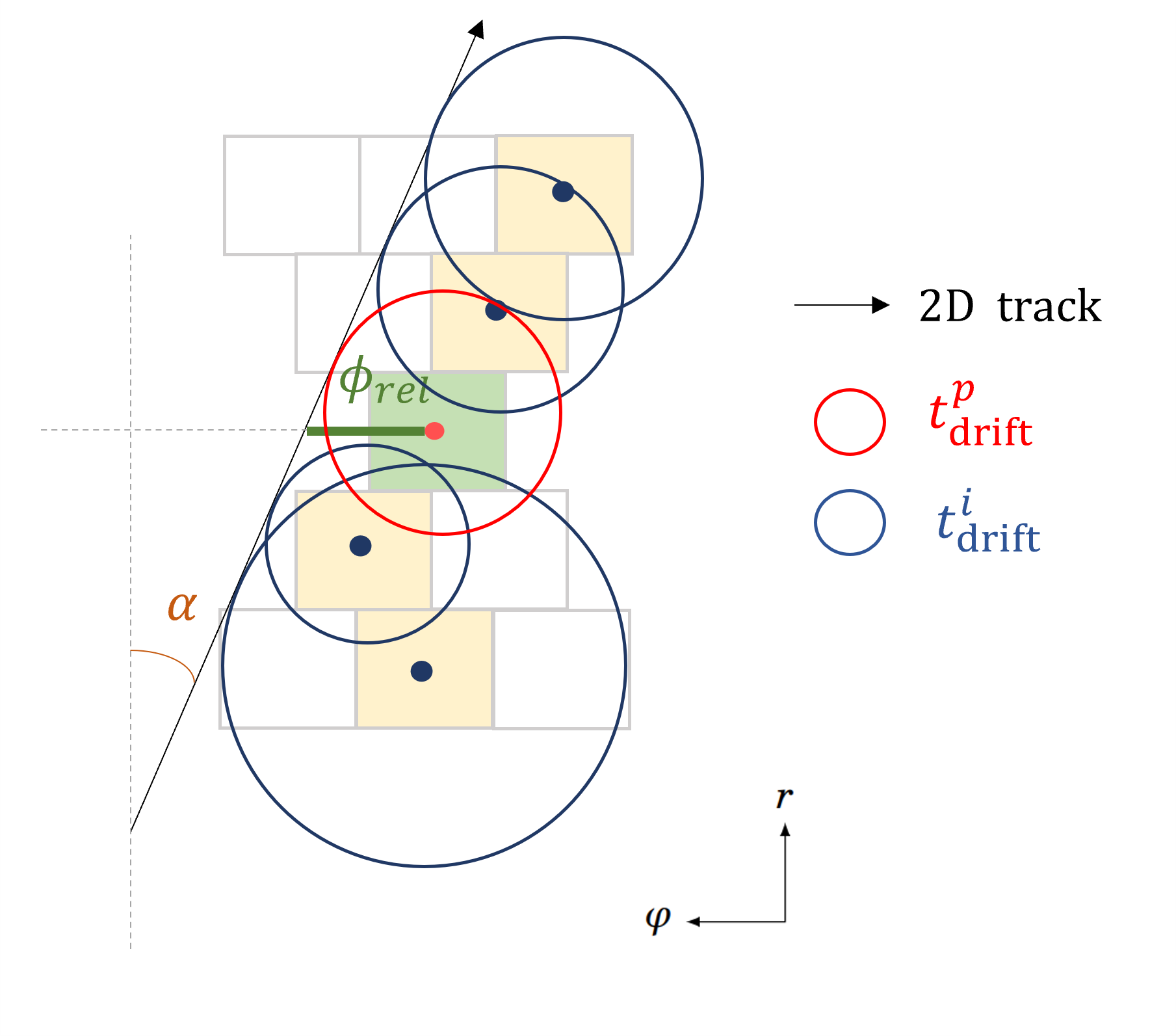}
\caption{Input variables of DNN track trigger from each TS, 
including 
$\phi_{rel}$, 
signed priority drift time $ t^p_{drift}$ and cross angle $\alpha$ for the priority wire, 
and 
extra drift time $t^{i}_{drift}$ for extra wires.}\label{fig:inputs}
\end{figure}
The DNN architecture is designed and optimized based on \texttt{PyTorch}~\cite{pytorch} and the Belle II analysis software (\texttt{basf2})~\cite{basf2,basf2_git} simulation with real CDC data and fully reconstructed tracks collected during Belle II operation. 
The DNN track trigger collects inputs from each selected TS across the nine SLs. For each TS, we extract the following features: the relative azimuthal angle $\phi_{rel} \equiv \phi_{cross} - \phi_B$, the signed priority drift time $\pm t^p_{drift}$, and the cross angle $\alpha \equiv \mu/(2r)$ for the priority wire. For stereo TSs, we also include the drift time $t^{i}_{drift}$ for extra wires, as shown in Fig.~\ref{fig:inputs}. Each input is scaled to the interval $(-1,1)$ to prevent bias among features and to normalize the model. Additionally, since not every wire in a TS registers a hit, $t^{i}_{drift}$ is scaled to $(0,1)$ for a valid hit and set to $-1$ for an invalid hit, enabling the DNN to learn the TS pattern. In total, there are 14 input features per stereo SL and 3 input features per axial SL, resulting in 71 inputs overall.

\begin{figure}[t]
\centering
\includegraphics[width=0.45\textwidth]{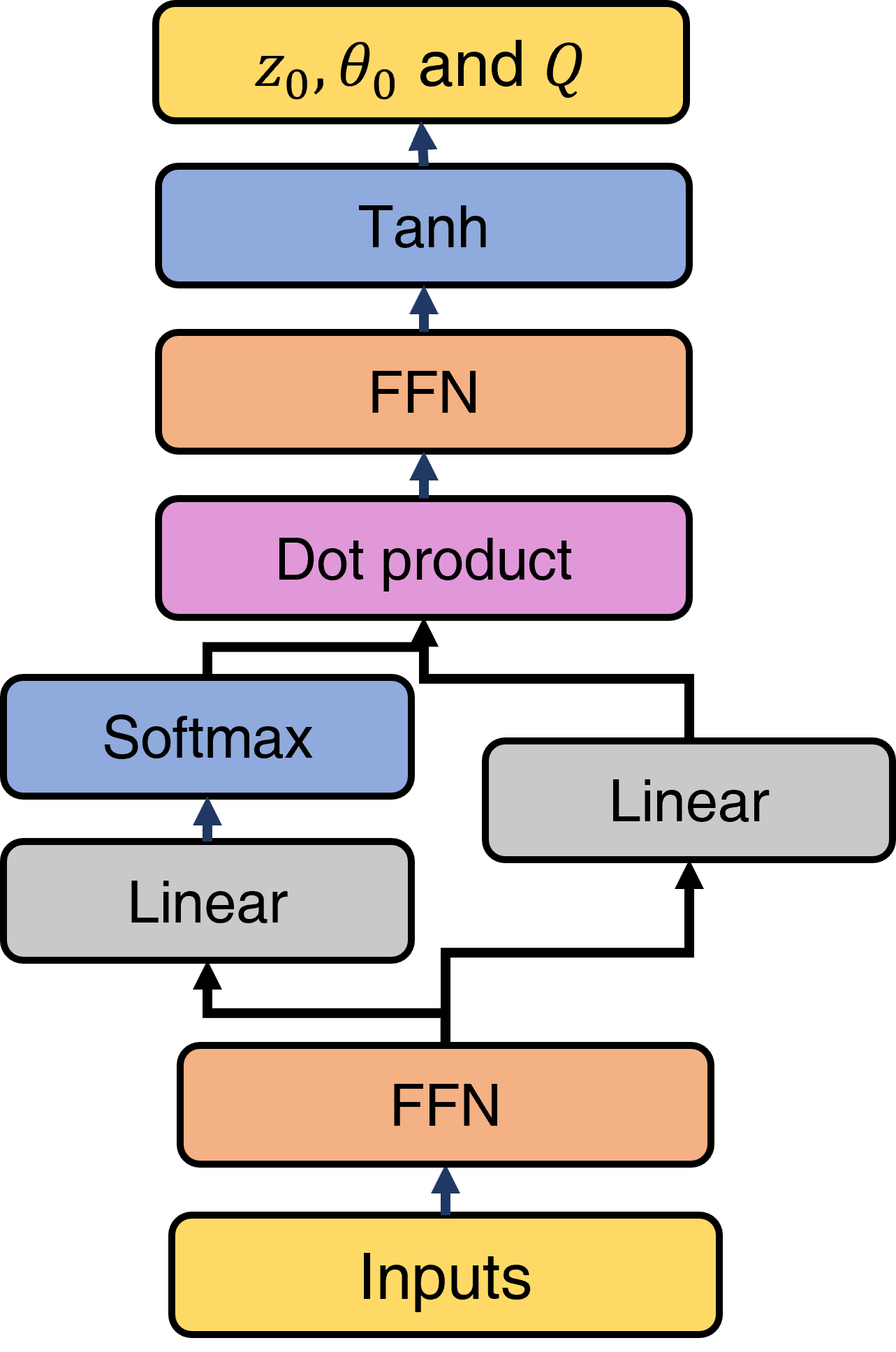}
\caption{Schematic of the designed DNN architecture.}
\label{fig:DNN}
\end{figure}
To handle the increasing number of background hits, we designed the DNN architecture as illustrated in Fig.~\ref{fig:DNN}. The inputs are first processed with a feed-forward network (FFN) for embedding. The FFN consists of multiple fully connected linear layers and LeakyReLU activation in between. A simplified self-attention block is applied for embedded feature selection. After that, another FFN is applied to predict the tracks' parameters and categories. Finally, we use Tanh to scale the output to $(-1,1)$. The simplified self-attention block works as follows:

\begin{align}
x_A = \text{Softmax}(xW_w) \cdot (xW_v + b_v),
\end{align}
where $x$ represents the embedded features, and $W_w$, $W_v$, and $b_v$ are trainable weight and bias matrices. This dot product operation enables the network to select the most relevant features. Compared to the original attention mechanism~\cite{vaswani2017attention}, instead of using three Query-Key-Value matrices and performing dot-product operations of query and key to get the attention weights, we only use one matrix in the algorithm. This is a compromise due to our FPGA resource and latency limitations. Subsequently, we perform track parameter prediction as follows:
\begin{align}
\begin{pmatrix}
    z_0\\[1ex]
    \theta_0\\[1ex]
    Q\\
\end{pmatrix}
= \tanh\bigl(\text{FFN}(x_A)\bigr), 
\end{align}

\noindent where
$z_0$ and $\theta_0$ are the track parameters, and $Q$ is an additional output that predicts whether the track is a signal track or background track, including fake tracks and real tracks outside the interaction point. This extra output is crucial for cases with high instantaneous luminosity on the order of $ 10^{34}~\text{cm}^{-2}\,\text{s}^{-1}$, where only a limited number of valid TSs may be available for accurate track parameter prediction due to the large number of background TSFs.

The model was implemented using \texttt{PyTorch}~\cite{pytorch} and trained in a supervised manner with target values for $z_0$, $\theta_0$, and $Q$ obtained from offline track reconstruction of real data collected during 2022. We employed a standard mean squared error (MSE) loss function taking equal contribution from three outputs and the \texttt{Adam} optimizer~\cite{Adam}.

Training hyperparameters and model architecture parameters, including learning rate, batch size, and the number of layers and the number of nodes per layer for each FFN, were tuned under the firmware limitations of the target FPGA (Virtex UltraScale XCVU160). The objective is to maximize the area under the ROC curve (AUC) for single-charge track prediction, which is made by applying cuts on both the $Q$ and $z_0$ outputs. The grid search was performed using the \texttt{Optuna} framework~\cite{optuna}. Considering the maximum available DSP units (1560) and the 
pipeline requirement of the L1 trigger system, the theoretical maximum number of multiply-accumulate operations (MAC) is limited to $4 \times 1560$, assuming one MAC per DSP and not using LUT for MACs. Table~\ref{tab:1} summarizes the tuning ranges and the resulting optimal parameters.

\begin{table}[t]
\centering

\begin{tabular}{l | c | c}
\hline
Parameters & Tuning Range & Optimal Value\\
\hline
Number of nodes per Layer for first FFN & (10, 50)  &27\\
Number of layers for first FFN          & (1, 2)    &1 \\
Number of nodes per Layer for second FFN& (10, 50)  &27\\
Number of layers for second FFN         & (1, 3)    &1 \\

\hline
\end{tabular}
\caption{Model architecture parameter tuning ranges and optimal values. We chose the best overall combination within the MAC limitations.}
\label{tab:1}
\end{table}
The model was 
initially 
trained using Belle II data from 2022, collected at a peak luminosity of $3.49\times10^{34}\,\text{cm}^{-2}\,\text{s}^{-1}$. The true track parameters were obtained using \texttt{basf2}. Tracks with $|z_0| < 1\,\text{cm}$ are considered signal (label: -1), while other tracks are treated as background (label: 1). In total, 3 million charged tracks were used for training, with a signal-to-background (S/N) ratio of approximately 2:1. A randomly weighted sampler was employed to balance the S/N ratio. To further enhance performance, five independent DNNs were trained to accommodate different cases of missing SL inputs. During the DNN commissioning process in 2024, additional 1 million charged tracks were collected, and fine-tuning was performed to adapt the model to the new background conditions.

To reduce resource consumption during firmware implementation, quantization was applied post-training. Using \texttt{neural-compressor}~\cite{neuralcompress}, the weights for each node were quantized according to:

\begin{align}
q &= \text{floor}\left(\frac{r}{s} + z\right), 
\end{align}

\noindent where "$\text{floor}$" is the floor rounding, $r$ represents the original weight in \texttt{float32}, $q$ denotes the quantized weight in \texttt{Int8}, and $s$ and $z$ are the scale factor and zero point, respectively, in \texttt{float32}. The scale factor and zero point for each node are tuned to achieve the best AUC. Once a weight is quantized to zero, we prune it. Taking an average over five experts, we have pruned 13\% of the weights. All inner nodes are quantized into a 16-bit signed fixed-point value, with 6 bits for the integer part (including the sign) and 10 bits for the fractional part. The outputs are quantized into a 13-bit signed fixed-point value, with 1 bit for the integer part (including the sign) and 12 bits for the fractional part.

\section{Firmware Implementation} \label{sec:firm}
The optimized DNN track trigger has been implemented on an AMD Virtex UltraScale FPGA (XCVU160) and 
meets several design requirements outlined below. To satisfy the pipeline constraints, 
the DNN must have larger throughput than four system clock cycles, corresponding to $127.216\,\text{MHz}/4 = 31.804\,\text{MHz}$, which is the frequency of data input from CDC. Additionally, the 3D track module's latency must be below 850\,ns to comply with the L1 trigger system's timing requirements. For the baseline module with UT3, the optical I/O introduces a latency of 515\,ns, leaving a maximum of 335\,ns for the remaining logic, including the neural network itself. In the case of the DNN implementation, the upgraded 25\,Gbps bandwidth reduces the I/O latency to 226\,ns, allowing the remaining logic to extend up to 624\,ns — equivalent to 80 system clock cycles.

\begin{figure}[t]
\centering
\includegraphics[width=0.89\textwidth]{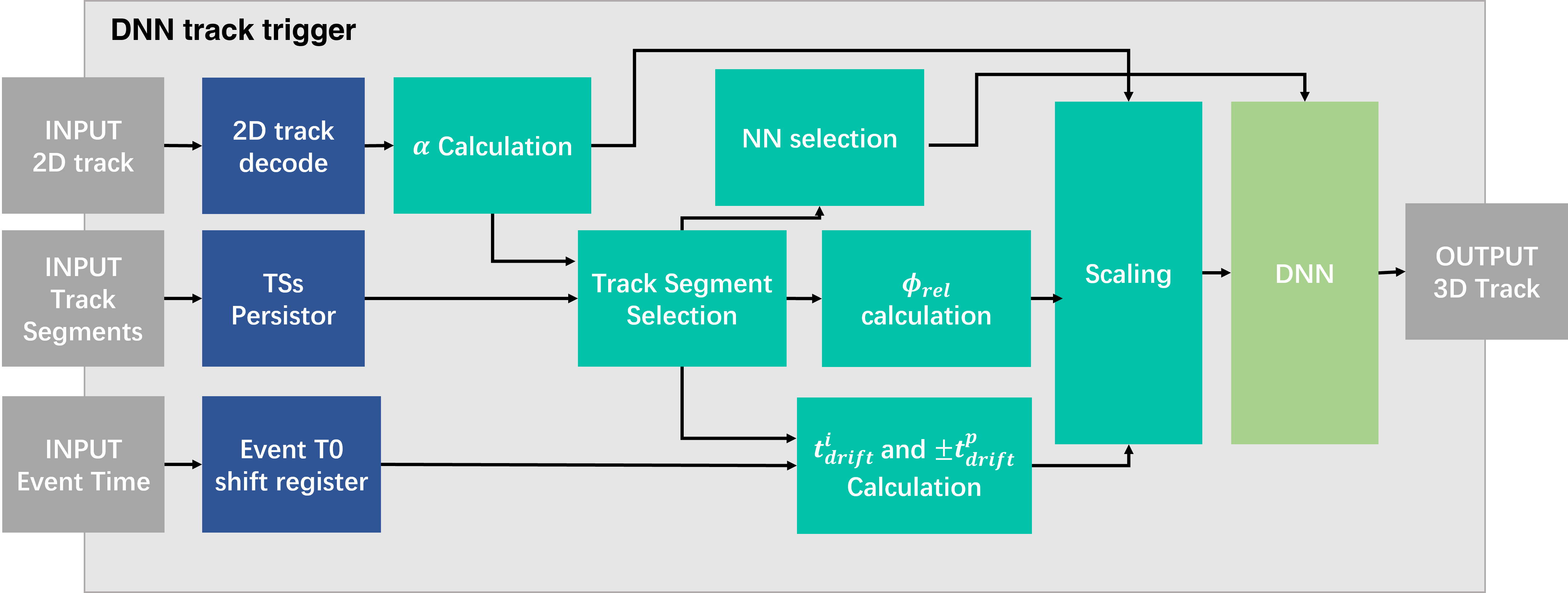}
\caption{Schematic diagram of the DNN track trigger firmware.}\label{fig:firmware}
\end{figure}

The designed firmware architecture is illustrated in Fig.~\ref{fig:firmware}. Due to drift-time-induced delays between TSs, and 2D tracking latency, the input TSs are stored in a FIFO for 27 system clock cycles to align them with the 2D tracks. Once a valid 2D track is detected, the system initiates preprocessing to collect all stored TSs and the event timing. This preprocessing stage performs track finding, input calculation, and scaling, and generates an enable signal for the DNN when a valid track is confirmed.

The DNN Intellectual Property (IP) core is generated using AMD Vitis HLS, which facilitates efficient architectural and weight modifications. The primary resource consumption stems from the multiply-accumulate operations (MACs) in the DNN, with a total of \[
(27\cdot4 + 71 + 3 + 2 + 1) \cdot 27 = 4995 \quad \text{MACs}.
\]
Although the FPGA’s DSP units are well-suited for these operations, the XCVU160 part provides only 1560 DSPs. Thus, we first divide the inputs for each linear layer into 4 groups, with each group processed in one clock cycle by reusing each DSP four times. Additionally, some MAC operations are offloaded to the LUTs. We specified floor-planning constraints for the logic cells of the linear layers to reduce routing complexity, assigning different ratios of LUT-based and DSP-based MACs for different layers; in general, approximately 35\% of MACs are implemented using the LUTs. The LeakyReLU activation and dot product operations are implemented directly with DSPs, while the nonlinear functions (softmax and tanh) are approximated using precomputed LUTs generated by the \texttt{hls4ml} library~\cite{hls4ml}.
Table~\ref{tab:2} summarizes the resource consumption, maximum frequency, and latency for the implemented DNN track trigger firmware.

\begin{table}[t]
\centering
\begin{tabular}{|c | c| c| c| c| c|}
\hline
FF & Distributed RAM & LUT & DSP & Maximum Frequency & Latency\\
\hline
12\% & 9\% & 53\% & 69\% & 127.216\,MHz & 593\,ns\\ 
\hline
\end{tabular}
\caption{Resource usage of the implemented DNN track trigger firmware.}
\label{tab:2}
\end{table}

\section{Performance} \label{sec:perf}
In this section, we evaluate the performance of the DNN track trigger and compare it with the baseline model using the experimental data collected in December 2024 during Belle II operation. During this period, DNN trigger did not join the trigger decision but only monitored the data. The DNN track trigger model used is trained with Belle II data collected in 2022 and fine-tuned using data from November 2024. The baseline model is trained with 2022 Belle II data.

Data for performance evaluation were specially taken to record both signal and background events. We perform full track reconstruction using \texttt{basf2}~\cite{basf2} with offline data to evaluate the trigger tracking performance. Trigger tracks produced by the DNN track trigger and baseline model were matched to the tracks from the full reconstruction by maximizing the number of shared CDC hits, with the additional requirement that at least 10\% of all CDC hits in a track were common between the two. Only matched tracks were used for the evaluation.

\begin{figure}[t]
\centering
\includegraphics[width=1\textwidth]{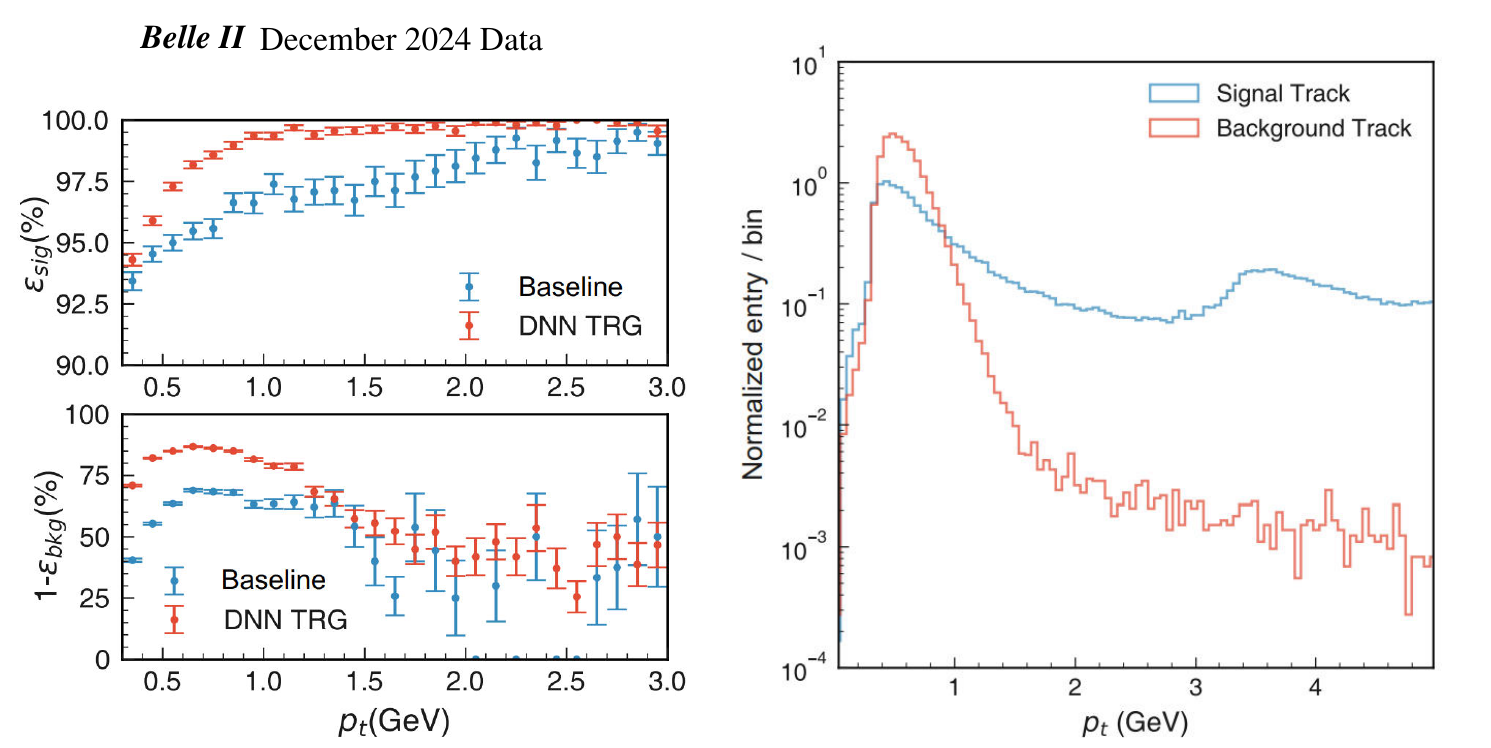}
\caption{Left: Signal efficiency ($\epsilon_{\text{sig}}$) and background rejection rate ($1-\epsilon_{\text{bkg}}$) as functions of $p_t$. Right: Normalized histograms of track $p_t$.}\label{fig:pts}
\end{figure}

We focus on two key metrics for the trigger performance: the signal track efficiency ($\epsilon_{\text{sig}}$) and the background track rejection rate ($1-\epsilon_{\text{bkg}}$), which are defined as
\begin{equation}
    \epsilon_{i} = \frac{N_{i,\text{pass}}}{N_{i,\text{total}}},
\end{equation}

\noindent where $i$ denotes the track type (signal or background), and $N_{i,\text{total}}$ and $N_{i,\text{pass}}$ represent the total number of tracks and the number of tracks passing the selection criteria, respectively. For the DNN track trigger, a combined selection is applied with $|z_0| < 50\,\text{cm}$ and $Q < 0.8$, chosen to minimize $\epsilon_{\text{bkg}}$ while maintaining the overall signal efficiency $\epsilon_{\text{sig}} > 98\%$ on the training sample. The baseline model employs a cut of $|z_0| < 15\,\text{cm}$ during trigger operation. Figure~\ref{fig:pts} shows the signal efficiency and background rejection rate as functions of track transverse momentum ($p_t$) for both models. 
Only tracks with $p_t>0.3$\,GeV, 
expected to go through every SL and form a valid track with trigger track-finding logic, 
were used in the analysis. 
Overall, the DNN track trigger improves $\epsilon_{\text{sig}}$ from 96\% to 98\% and the background rejection from 60\% to 83\%. Both DNN track trigger and the baseline model saw the $\epsilon_{\text{sig}}$ drop at $p_t <0.9$\,GeV. DNN track trigger achieved stable $\epsilon_{\text{sig}} >99\%$ at $p_t \geq0.9$\,GeV. For background tracks, the remaining dominant background track with $p_t\leq1.2$\,GeV can be halved using the DNN track trigger.

\begin{figure}[t]
\centering
\includegraphics[width=0.89\textwidth]{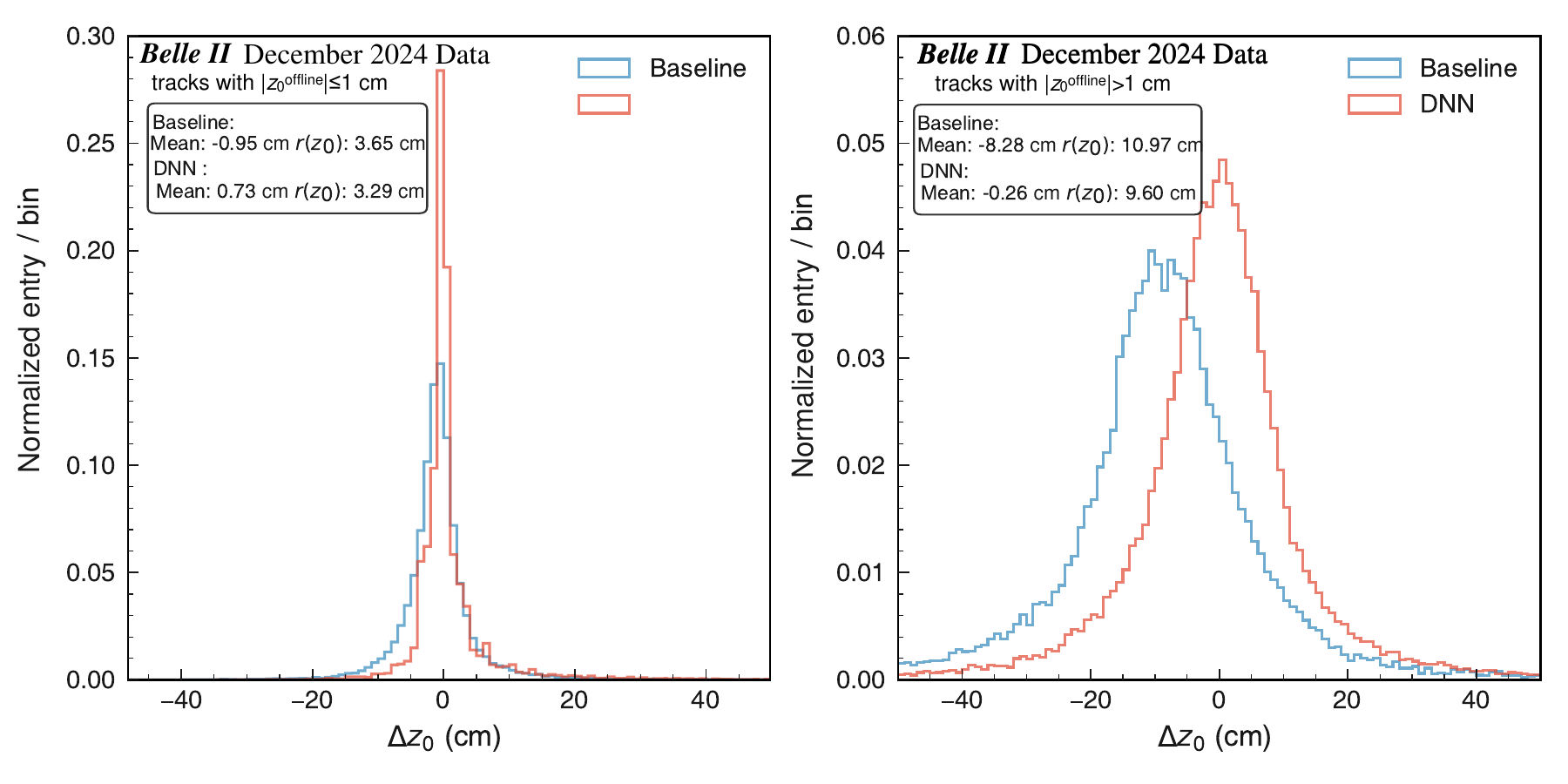}
\caption{Normalized histograms of $\Delta z_0 \equiv z_0^{\text{trg}} - z_0^{\text{offline}}$ distributions for baseline and DNN track trigger. Left: signal tracks with $|z_0^{\text{offline}}| <= 1\,\text{cm}$. Right: background tracks ($|z_0^{\text{offline}}| > 1\,\text{cm}$). Each histogram is normalized to unit area for comparison.}\label{fig:dz0}.
\end{figure}

We also evaluate the resolutions of the predicted track parameters. Due to different tracking efficiencies, only offline tracks that have been found in both DNN track trigger and baseline model were used for analysis. The $z_0$ resolution comparison is shown in Fig.~\ref{fig:dz0}. We define the resolution for track parameter $i$ as:

\begin{equation}
    r({i}) = \text{std}(\Delta i\in[P_{2.5},P_{97.5}]),
\end{equation}

\noindent where $i$ is the track parameter, $\Delta i$ is the distribution of $i^\text{trg} - i^\text{offline}$, $P_q$ is the q-th quantile of the distribution of $\Delta i$ and $\text{std}$ is the standard deviation. For both the signal and background track cases, we have improved the $r(z_0)$ by 8\% on average. And for the background track case, the mean value shift, which was a known issue with the baseline model leading to more background tracks 
misclassified as signal, 
is well addressed with the DNN track trigger. 

\begin{figure}[t]
\centering
\includegraphics[width=0.89\textwidth]{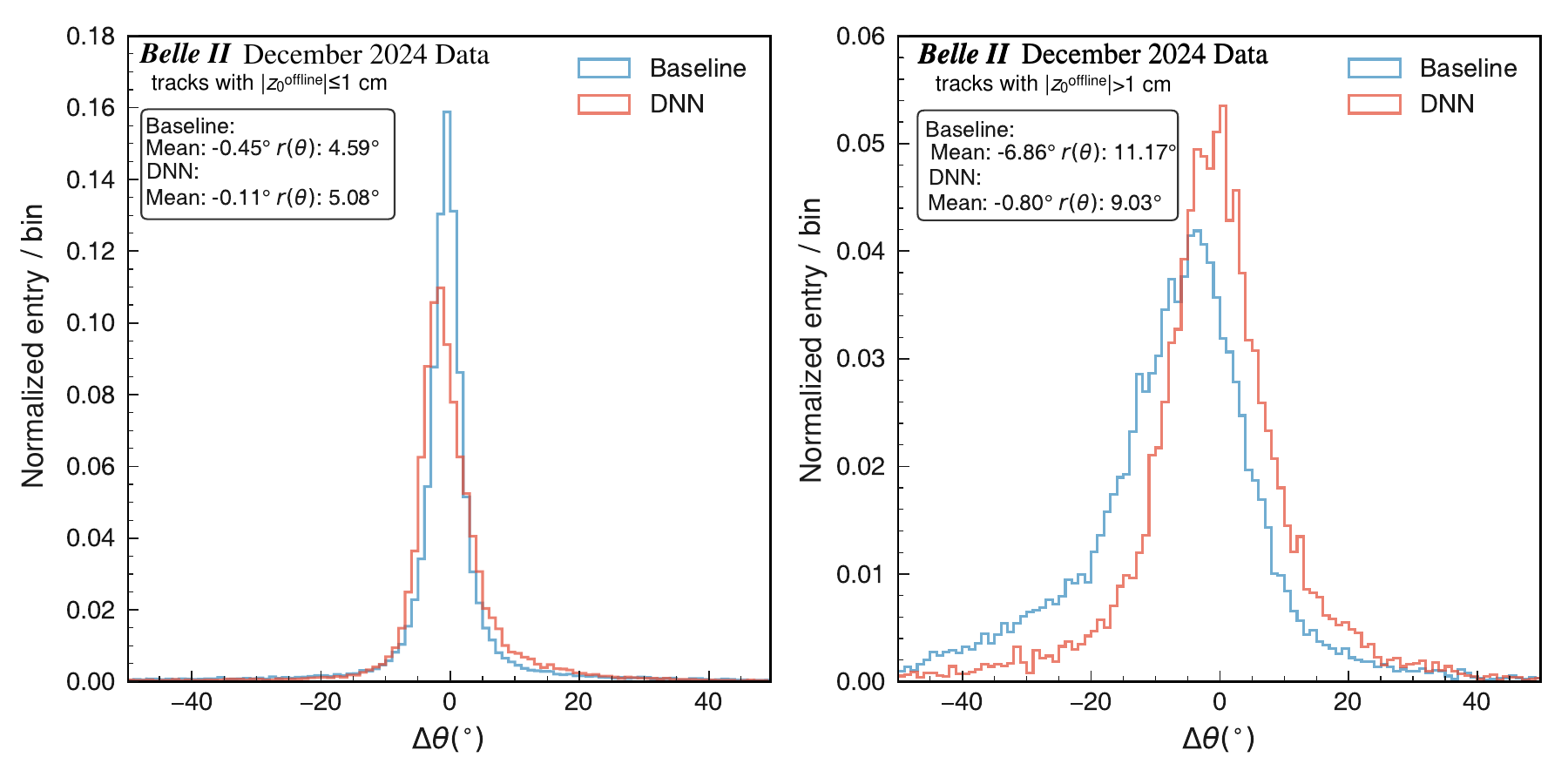}
\caption{Normalized histograms of $\Delta \theta \equiv \theta_0^{\text{trg}} - \theta_0^{\text{offline}}$ distributions for the baseline and DNN track trigger. Left: signal tracks. Right: background tracks. Each histogram is normalized to unit area for comparison.}\label{fig:dtheta}
\end{figure}

The impact of the DNN track trigger on $\theta$ is demonstrated 
in Fig.~\ref{fig:dtheta}. 
For signal tracks, 
we observe a degradation in $\theta$ resolution 
also with a small peak shift. 
In contrast, with the DNN track trigger 
we obtain a better background track $\theta$ distribution 
with no peak shift compared with the baseline model. 
However, since the current trigger logic does not use $\theta$ 
for either track matching or trigger decision~\cite{GRL}, 
these $\theta$ effects are irrelevant for the trigger performance.

\begin{figure}[t]
\centering
\includegraphics[width=0.6\textwidth]{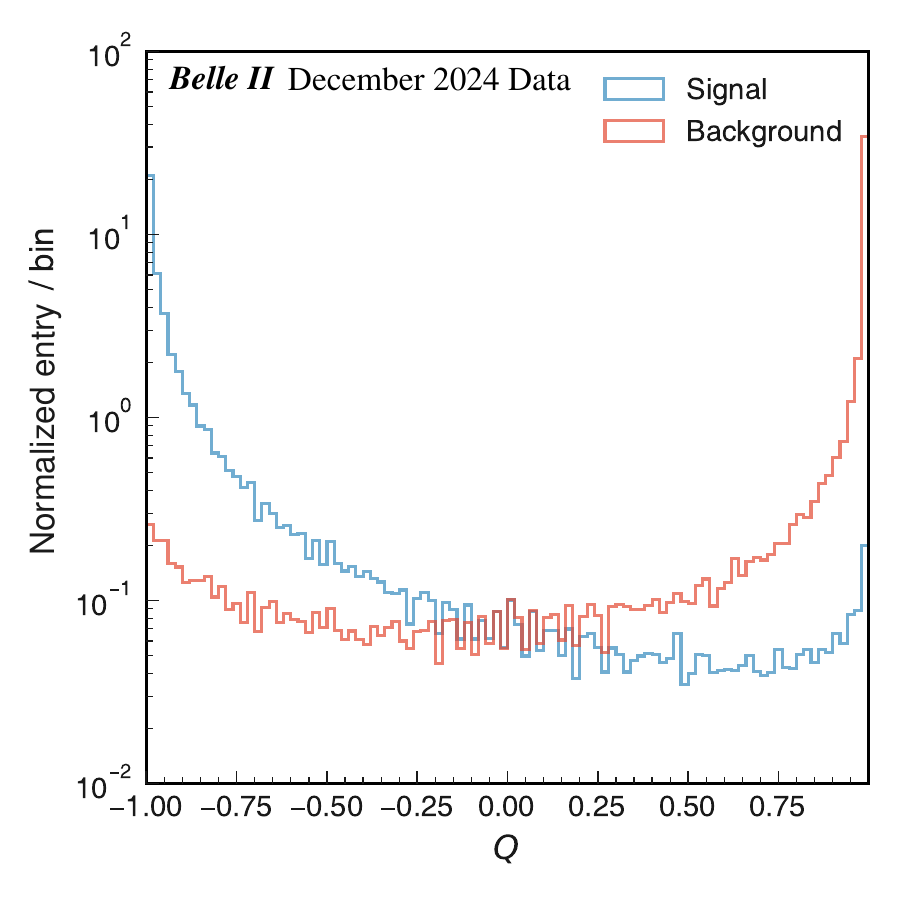}
\caption{Normalized histograms of the $Q$ output from the DNN track trigger.}\label{fig:Q}
\end{figure}

The DNN track trigger $Q$ outputs are shown in Fig.~\ref{fig:Q}. With the $Q$, the DNN track trigger demonstrates an accuracy of 93\% for track classification. 

We have monitored the track trigger rate, which is defined as the rate of 3D trigger track satisfying the selection criteria. 
The same described above selection criteria are used for the baseline and DNN track trigger. 
During Belle~II operation in December 2024, with an average instantaneous luminosity of $2.75 \times 10^{34}~\text{cm}^{-2}\,\text{s}^{-1}$, the average track trigger rate was reduced from 4.32\,kHz to 2.68\,kHz through the application of the DNN track trigger.

%

\section{Conclusion} \label{sec:summary}
In this work, we have developed a Deep Neural Network (DNN) track trigger for the Belle II experiment to achieve robust track fitting and classification 
against high beam-induced background 
at the hardware trigger level. The implementation is deployed on the UT4 board with an AMD Virtex UltraScale FPGA using high-level synthesis techniques. The DNN track trigger processes inputs from two-dimensional tracks and track segments, which contain both stereo and axial wires from the Central Drift Chamber (CDC). By leveraging drift time information and hit patterns from each wire in the track segments and using a DNN with a simplified self-attention architecture, the DNN track trigger demonstrates a significant reduction in the total track trigger rate 
by 37\% while maintaining higher efficiency 
for signal tracks across all transverse momentum regions compared to the existing MLP-based track trigger. This improvement ensures that the trigger rate remains within the limitations of the Belle II data acquisition system as the experiment moves toward higher luminosity operation. This is the first implementation of an attention-based DNN in the hardware trigger system for collider experiments. The DNN track trigger is expected to be used in Belle~II starting from 2025 data-taking. 

This work was supported by JSPS KAKENHI Grant Number JP23H05433  and JP22K21347.

\bibliographystyle{elsarticle-num} \bibliography{biblograph}

\end{document}